\documentclass{svproc}

\usepackage{url}

\usepackage{amsmath}
\usepackage{latexsym}
\usepackage{amsfonts}
\usepackage{amssymb}
\usepackage{booktabs}
\usepackage[free-standing-units=true]{siunitx}
\usepackage{cite}
\usepackage{xcolor}

\usepackage[misc]{ifsym} 

\usepackage{times}	

\usepackage[pdftex]{graphicx} 
\DeclareGraphicsExtensions{.pdf}

\usepackage{paralist}	
\usepackage{enumitem}

\usepackage{xspace}

\newcommand{\eg}{e.g.,\xspace}
\newcommand{\ie}{i.e.,\xspace}

\usepackage{hyperref}
\hypersetup{
    unicode=false,		
    colorlinks=true,	
    linkcolor=black,	
    citecolor=black,	
    filecolor=black,   	
    urlcolor=black		
}

\usepackage{fancyhdr}
\usepackage{ifthen}

\newcommand{\prob}{\mathrm{Pr}}

\usepackage{mathtools} 
\DeclarePairedDelimiter\abs{\lvert}{\rvert}

\newcommand{\actualCTE}{\mathrm{CTE}_a}

\newcommand{\estimatedCTE}{\mathrm{CTE}_e}
\newcommand{\estimatedHE}{\mathrm{HE}_e}
\newcommand{\offset}{\mathtt{offset}}
\newcommand{\assuredTaxi}{\mathtt{Assured Taxi}}

\usepackage{acronym}

\acrodef{ac}[AC]{assurance case}

\acrodef{cdf}[CDF]{cumulative distribution function}
\acrodef{cms}[CMS]{Contingency Management System}
\acrodef{cnn}[CNN]{convolutional neural network}
\acrodef{cte}[CTE]{cross track error}
\acrodef{darpa}[DARPA]{Defense Advanced Research Projects Agency}
\acrodef{dac}[DAC]{dynamic assurance case}
\acrodef{dag}[DAG]{Directed Acyclic Graph}
\acrodef{dbn}[DBN]{dynamic Bayesian network}
\acrodef{dnn}[DNN]{deep neural network}
\acrodef{dsc}[DSC]{dynamic safety case}
\acrodef{dsl}[DSL]{domain-specific language}
\acrodef{gp}[GP]{Gaussian process}
\acrodefplural{gp}[GPs]{Gaussian processes} 
\acrodef{lec}[LEC]{learning-enabled component}
\acrodef{les}[LES]{learning-enabled system}
\acrodef{ood}[OOD]{out of distribution}
\acrodef{pdf}[PDF]{probability density function}
\acrodef{pid}[PID]{proportional-integral-derivative}
\acrodef{rl}[RL]{reinforcement learning}
\acrodef{rv}[RV]{random variable}
\acrodef{uas}[UAS]{unmanned aircraft system} 
\acrodef{uq}[UQ]{uncertainty quantification}
\acrodef{bn}[BN]{Bayesian Network}

\acrodef{cpd}[CPD]{conditional probability distribution}
\acrodef{cpt}[CPT]{Conditional Probability Table}
\acrodef{dsd}[DSD]{Dataset Shift Detector}
\acrodef{mcte}[MCTE]{Measured Cross Track Error}
\acrodef{ml}[ML]{machine learning}
\acrodef{he}[HE]{heading error}
\acrodef{ann}[ANN]{artificial neural network}
\acrodef{map}[MAP]{maximum a posteriori probability}
\acrodef{mcmc}[MCMC]{Markov Chain Monte Carlo}

\begin{document}
\mainmatter             

\title{Quantifying Assurance in Learning-enabled Systems\thanks{This work was
supported by the Defense Advanced Research Projects Agency (DARPA) and the Air 
Force Research Laboratory (AFRL) under contract FA8750-18-C-0094 of the Assured 
Autonomy Program. The opinions, findings, recommendations or conclusions 
expressed are those of the authors and should not be interpreted as representing 
the official views or policies of DARPA, AFRL, the Department of Defense, or 
the United States Government.}}

\titlerunning{Quantifying Assurance in Learning-enabled Systems}  

\author{Erfan Asaadi \and Ewen Denney \and 
					Ganesh Pai$^{(\textrm{\Letter})}$}
\authorrunning{E.~Asaadi, E.~Denney, and G.~Pai} 

\tocauthor{Erfan Asaadi, Ewen Denney, and Ganesh Pai}
\institute{KBR, Inc., NASA Research Park\\
			Moffett Field, CA 94035, USA\\
\email{\{easaadi, edenney, gpai\}@sgt-inc.com}}
\maketitle
\thispagestyle{fancy}           

\begin{abstract}

\acused{les}

Dependability assurance of systems embedding \ac{ml} components---so called 
\emph{\aclp{les}} (\acsp{les})---is a key step for their use in safety-critical 
applications. In emerging standardization and guidance efforts, there is 
a growing consensus in the value of using assurance cases for that purpose. 
This paper develops a quantitative notion of assurance that an \ac{les} is 
dependable, as a core component of its assurance case, also extending our prior 
work that applied to \ac{ml} \emph{components}. 
Specifically, we characterize \ac{les} assurance in the form of 
\emph{assurance measures}: a probabilistic quantification of confidence that 
an \ac{les} possesses system-level properties associated with functional capabilities and dependability attributes. We illustrate the utility of 
assurance measures by application to a real world autonomous aviation system, 
also describing their role both in 
\begin{inparaenum}[\itshape i\upshape)]
	\item guiding high-level, runtime risk mitigation decisions and 
	\item as a core component of the associated \emph{dynamic assurance case}.
\end{inparaenum}
\\


\end{abstract}

\section{Introduction}\label{s:intro}

\acused{ac}

The pursuit of developing systems with increasingly autonomous 
capabilities is amongst the main reasons for the emergence of 
\emph{\aclp{les}} (\acsp{les}), \ie systems embedding \acf{ml} 
based software components. 
%
%
There is a growing consensus in autonomy standardization efforts~\cite{ul4600std} 
on the value of using \emph{\aclp{ac}} (\acsp{ac}) as the mechanism by which to convince various stakeholders that an \ac{les} can be relied upon. \acp{ac} have 
been successfully used for safety assurance of novel aviation applications 
where---like \acp{les}---regulations and standards continue to be under 
development~\cite{cdp-atio-2017}.
However, \acp{les} pose particular assurance challenges~\cite{mcdermid2019}
and existing \ac{ac} technologies may not be sufficient, requiring
a framework where the system and its \ac{ac} evolve in tandem~\cite{dhp-icse2015}. 
Here too, there are specific additional challenges:
first, structured arguments\footnote{The systematic reasoning that captures 
the rationale why specific conclusions, \eg of system safety, can be drawn 
from the evidence supplied.} in many \acp{ac} are effectively \emph{static}, 
\ie they are usually developed prior to system deployment under assumptions 
about the environment and intended system behavior. 
Evolution of the system or its \ac{ml} components (\eg via online learning, 
or by adaptation in operation) can render invalid a previously accepted 
\ac{ac}. 
In principle, although it is possible to dynamically evolve structured 
arguments~\cite{dhp-icse2015}, since their role is primarily to convince 
human stakeholders, it makes more sense for such updates to happen between 
missions at well-defined points. 

Second, an operational evaluation of the extent of assurance in an \ac{les} 
(or its \ac{ml} components, where appropriate) is a valuable system-level 
indicator of continued fitness for purpose. That, in turn, can facilitate 
potential intervention and counter-measures when assurance drops below 
an acceptable level during a mission. Indeed, \emph{online assurance updates} 
that are aimed at machine consumption must necessarily be in a computable 
form, \eg using a formal language, such as a logic, or as a quantification. So 
far as we are aware, prevailing notions of \acp{ac} do not yet admit such 
evaluation. 
Prior efforts at \ac{ac} confidence assessment~\cite{dhp-esem-11, guiochet2019} 
have focused on the argument structure rather than the system itself, and face 
challenges in repeatable, objective validation due to their reliance on 
subjective data. They have also not been applied to \acp{les}. 
Thus, there is a general need to capture a computable form of assurance 
to bolster an otherwise qualitative \ac{ac}. Note that although a qualitative 
\ac{ac} may well refer to quantitative evidence items, here we are identifying 
the necessity to have quantified assurance as a core facet of \ac{les} \acp{ac}. 

This paper focuses on the problem of assurance quantification, deferring its 
use in dynamic updates to future work. The main contribution is an 
approach to characterize assurance in an \ac{les} through \ac{uq} of 
system-level dependability attributes, demonstrated by application to an
aviation domain \ac{les}.

\acused{cte}

\section{Methodology}\label{s:approach}
 
Previously~\cite{adp-edcc2019}, we have described how assurance of \ac{ml} 
\emph{components} in an \ac{les} can be characterized through \ac{uq} of
component-level properties associated with the corresponding (component-level)
dependability attributes. Here, we extend our methodology to the system-level, 
relying on the following concepts:
\emph{assurance} is the provision of (justified) confidence that an 
\emph{item}---\ie a (learning-enabled) component, system, or service---possesses 
the relevant assurance properties. An \emph{assurance property} is a logical, 
possibly probabilistic characteristic associated with \emph{dependability 
attributes}~\cite{dependability-taxonomy} and functional capabilities. One or 
more assurance properties applied to a particular item give an \emph{assurance 
claim}\footnote{Henceforth, we do not distinguish assurance properties from 
assurance claims.}. 
An \emph{assurance measure} characterizes the extent of confidence 
that an assurance property holds for an item through a probabilistic 
quantification of uncertainty. It can be seen as implementing a \ac{uq} model 
on which to query the confidence in an assurance property.\footnote{When the 
assurance property is itself probabilistic, the corresponding assurance measure
is deterministic, \ie either 0 or 1.}

In general, we can define multiple assurance properties (and assurance measures), based on the \ac{les} functionality and dependability attributes for which 
assurance is sought. For example, the proposition ``\emph{the aircraft location 
does not exceed a specified lateral offset from the runway centerline during taxiing}'' is a system-level assurance claim associated with the attribute 
of \emph{reliability}. 
Similarly, the assurance property ``\emph{the aircraft does not veer off 
the sides of the runway during taxiing}'' is associated with the attribute 
of \emph{system safety}. Such assurance properties directly map to the claims 
made in the structured arguments of an \ac{les} \acl{ac}. Thus, we can leverage 
the methodology for creating structured arguments~\cite{dp-jase2018} to also 
specify assurance properties. 

For quantification, we mainly consider assurance measures for those 
system-level properties that can be reasonably and feasibly quantified.
For example, assurance measures for the preceding example quantify the 
uncertainty that the aircraft location does not exceed, respectively, the 
specified lateral offset from the runway centerline (reliability), and half 
the width of the runway pavement (safety), over the duration of taxiing. 

\acp{les} used in safety-critical applications, especially aviation, are 
effectively stochastic dynamical systems. The insights from this 
observation are that we can:
\begin{inparaenum}[\itshape i\upshape)]
	\item capture \ac{les} behavior through model-based representations 
	of the underlying stochastic process;
	\item view system-level assurance properties as specific realizations of 
	particular \acp{rv} of that process; and 
	\item express confidence in the assurance properties---\ie the assurance 
	measures---by propagating uncertainty through the model to determine
	the distributions over the corresponding \acp{rv}. 
\end{inparaenum} 

One challenge is selecting an appropriate model and representation of the
stochastic process to be used to model \acp{les}. Although there is not a 
generic answer for this, such a model could be built, for example, 
by eliciting the expected system behavior from domain experts, by transforming 
a formal system description, using model fitting and statistical optimization 
techniques applied to (pre-deployment) system simulation and execution traces, 
or through a combination of the three. For \acp{les}, a formal system description 
may be often unavailable. 
As such, we rely on elicitation and statistical techniques, using Bayesian 
models where possible, making allowance to admit and use other well-known, 
related stochastic process models---such as Markov chains---and leveraging 
data from analytical representations of system dynamics, simulations, and 
execution. 
The Bayesian concepts of \emph{credible} intervals and regions---determined 
on the posterior distribution of the \acp{rv} for assurance properties---give 
a formal footing to the intuitive, subjective notion of confidence that 
usually accompanies claims in assurance arguments, and \acp{ac} in 
general~\cite{Hawkins2011}.

\section{Illustrative Example -- Runway Centerline Tracking}\label{s:example}

\acused{cte}
\acused{he}

\subsection{System Description}

\begin{figure}[t]
	\centering
	\includegraphics[width=0.6\textwidth]{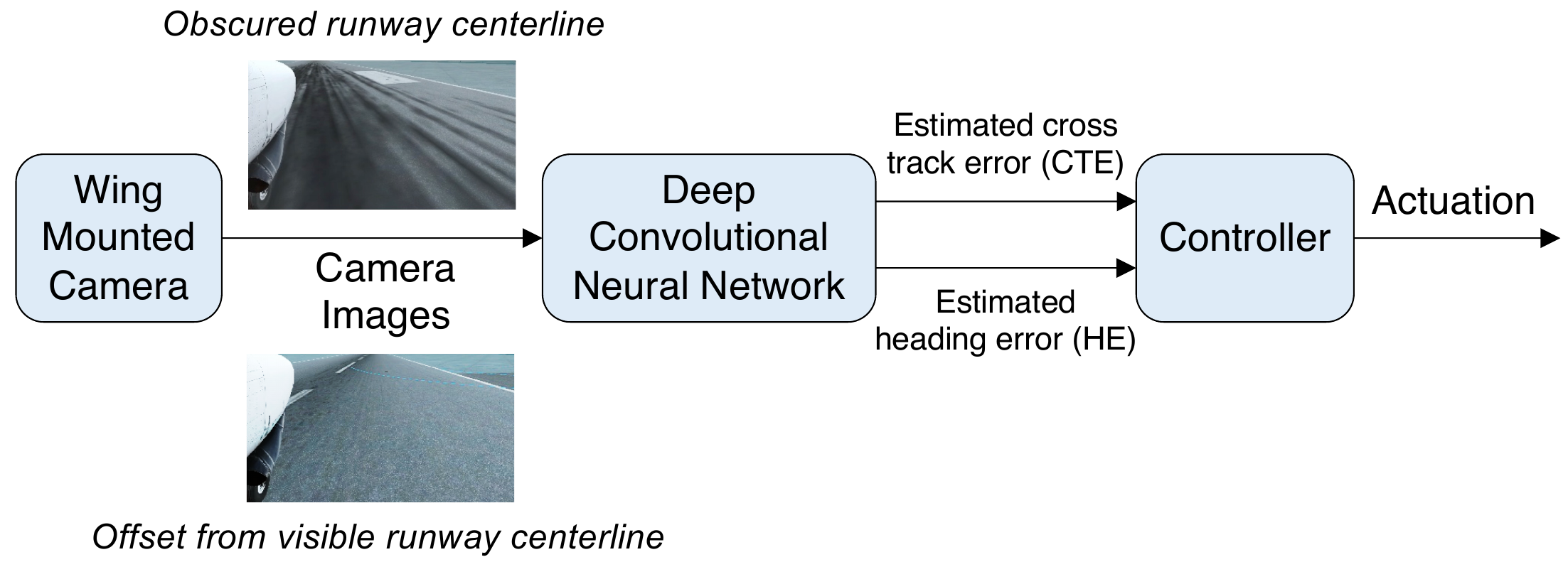}
	\caption{Pipeline architecture to implement an autonomous taxiing 
			capability in a \ac{uas}.}
	\label{f:uas-example}
\end{figure}

To show our methodology is feasible, we now apply it to quantify 
assurance in an aviation domain \ac{les} supplied by our industrial 
collaborators: a \ac{uas} embedding an \ac{ml} component, trained offline 
using supervised learning, to support an autonomous taxiing capability. The 
broader goal is to enable safe aircraft movement on a runway without human pilot input.  
Fig.~\ref{f:uas-example} shows a simplified \emph{pipeline architecture} used 
to realize this capability. A deep \ac{cnn} implements a perception function 
that ingests video images from a wing-mounted camera pointed to the nose of 
the aircraft.
The input layer is $(360 \times 200)$ pixels $\times~3$ channels wide; the 
network size and complexity is of the order of $100$ layers with greater than 
two million tunable parameters. Effectively, this \ac{ml} component performs
regression under supervised learning producing estimates of \emph{\acl{cte}}
(\acs{cte})\footnote{The horizontal distance between the aircraft nose wheel 
and the runway centerline.} and \emph{\acl{he}} (\acs{he})\footnote{\emph{Heading} 
refers to the compass direction in which an object is pointed; \acf{he} here, 
is thus the angular distance between the aircraft heading and the runway 
heading.} as output. These estimates are input to a classical \ac{pid} 
controller that generates the appropriate steering and actuation signals.

\subsection{Assurance Properties}\label{sss:uas-assurance-properties}

The main objective during taxiing (autonomously, or under pilot control) is
to safely follow the runway (or taxiway) centerline. Safety during taxiing 
entails avoiding \emph{lateral runway overrun}, \ie not veering off the sides 
of the runway pavement. Although avoiding obstacles on the runway is also a 
safety concern, it is a separate assurance property that we do not consider 
in this paper. 
Thus, safety can be achieved here, in part, by meeting a performance objective 
of maintaining an acceptable lateral offset (ideally zero) on either side of 
the runway centerline during a taxi \emph{mission} from starting taxi to 
stopping (or taking off).\footnote{Our industry collaborators elicited the 
exact performance objectives from current and proficient professional pilots.}
In other words, the closer the aircraft is to the runway centerline 
during taxiing, the less likely it is to veer off the sides of the runway. 

This performance objective relates to the attribute of reliability, where 
\emph{taxi failure} is considered to be the violation of the specified 
lateral offset. 
Here, we focus on the corresponding assurance property, 
$\assuredTaxi: \abs{\actualCTE} < \offset$, where $\offset=2\meter$ is the 
maximum acceptable lateral offset on either side of the runway centerline for 
this application and aircraft type. $\actualCTE$, which is the true (or actual) 
\ac{cte} for the \ac{uas}, is a signed, real valued scalar; the absolute value 
gives the magnitude of the offset, and the sign indicates where the \ac{uas} is 
located relative to the centerline, \ie to its left or its right.

\subsection{Assurance Quantification}\label{sss:uas-uq}

\acused{dbn}

\subsubsection{Model Choice} 

The assurance measure corresponding to $\assuredTaxi$, establishes 
$\prob\left(\abs{\actualCTE} < 2\meter\right)$, which characterizes the 
uncertainty (or conversely, confidence) in the true (or actual) \ac{cte} 
($\actualCTE$) relative to the specified offset. $\actualCTE$ evolves in 
time as the \ac{pid} controller responds to \emph{estimates} of \ac{cte} 
and \ac{he}, themselves the responses of the deep \ac{cnn} component, to 
runway images captured by the wing mounted camera (see Fig.~\ref{f:uas-example}).
$\actualCTE$ is thus uncertain and depends on other variables, of which those that 
can be observed are the estimated \ac{cte} ($\estimatedCTE$), estimated \ac{he} 
($\estimatedHE$), and a sequence of images. 
We can also model the controller behavior in terms of a time series evolution of 
$\actualCTE$ since, during taxiing, the true \ac{cte} at a given time is 
affected by the controller actuation signals at prior times. 

An abstracted model of \ac{les} behavior is reflected in the joint 
distribution of the relevant observed and uncertain variables. 
In fact, a \emph{\acl{dbn}} (\acs{dbn})~\cite{ml-book} is a convenient and 
compact representation of this joint distribution, as we will see subsequently in this section. It takes into account the temporal evolution of the variables 
and their (known or assumed) conditional independence relations. 
Thus, to determine the assurance measure, we effectively seek to quantify the 
(posterior) distribution over $\actualCTE$, given a sequence of runway 
images, the estimates of \ac{cte} and \ac{he} produced by the \ac{ml} component, 
and the controller behavior, as a query over the corresponding \ac{dbn} model.

\begin{table}[t]
	\centering
	\caption{\ac{dbn} model variables.}
	\label{t:uas-model-vars}
	\includegraphics[width=\columnwidth]{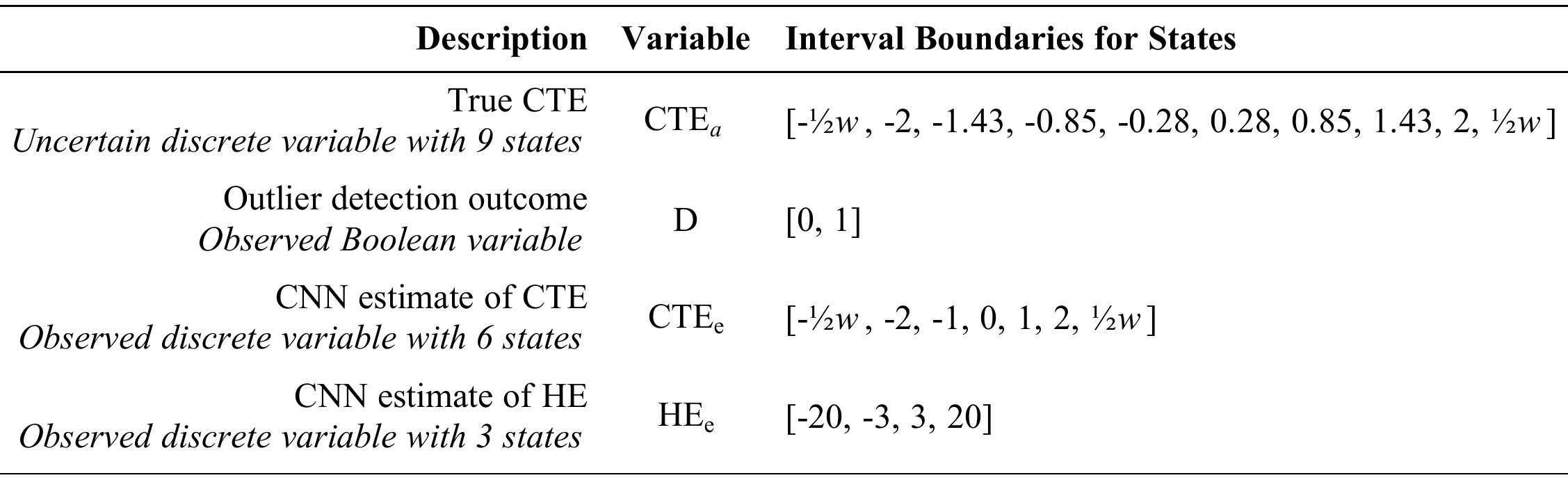}
\end{table}

\subsubsection{Model Variables}

Model variables can be discrete or continuous, and there are tradeoffs
between information loss and computational cost involved in the choice. 
Table~\ref{t:uas-model-vars} lists the discrete variables we have chosen, 
giving the interval boundaries for their states. The choice of the intervals 
that constitute the states of the variables has been based, in part, on:
\begin{inparaenum}[\itshape i\upshape)]
	\item domain knowledge, 
	\item an assessment of the data sampled from the environments 
	used for training and testing the \ac{cnn}, and 
	\item the need to develop an executable model that was modest in 
	its computational needs. 
\end{inparaenum}

Here, $w$ is the width of the runway in meters, and negative values 
represent \ac{cte} measured on the left of the runway centerline. The \ac{he} 
is given in degrees, while $D$ is dimensionless. An additional variable ($I$, 
not shown in Table~\ref{t:uas-model-vars}) models the runway image captured 
from the camera video feed as a vector of values in the range $[0\dots1]$ representing normalized pixel values. 
The Boolean variable $D$ represents the detection of outliers in camera 
image data. Such outliers may manifest due to various causes, including 
camera errors and \emph{covariate shift}, \ie when the data input to 
the \ac{cnn} has a distribution different from that of its training data. 
Note that the \ac{les} shown in Fig.~\ref{f:uas-example} does not 
indicate whether or not it includes a mechanism to detect outliers or 
covariate shift. However, we include this variable here, motivated by 
our earlier work on component-level assurance quantification of the 
\ac{cnn}~\cite{adp-edcc2019}, which revealed its susceptibility to outlier 
images. In fact, $D$ models a runtime monitor for detecting \ac{ood} inputs to 
the \ac{cnn}.

\begin{figure}[t]
	\centering
	\includegraphics[width=0.65\columnwidth]{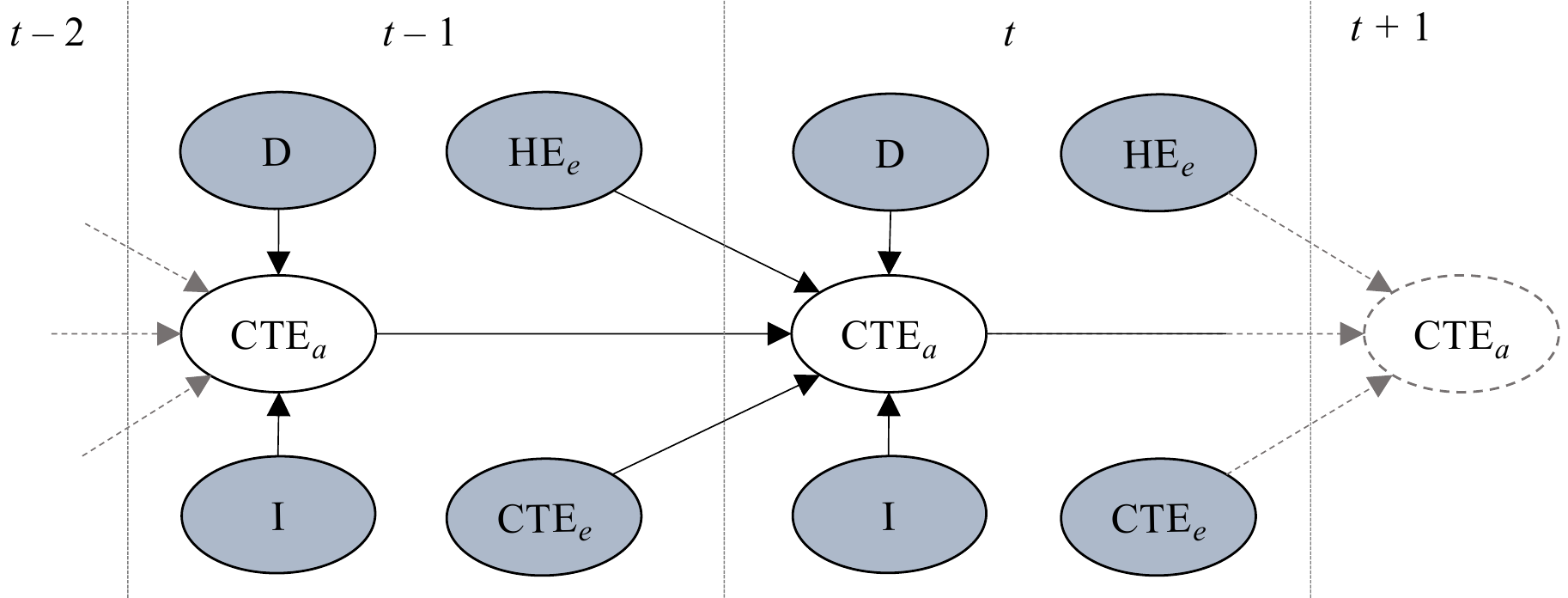}
	\caption{\ac{dbn} structure for assurance quantification, showing  
		two adjacent slices at times $t-1$, and $t$; shaded nodes represent
		observed variables, clear nodes are the uncertain, latent variables.}
	\label{f:uas-dbn}
\end{figure}

\subsubsection{Model Structure}

Each variable in Table~\ref{t:uas-model-vars} is indexed over time: we will 
denote a variable $X$ at time $t$ as $X^{(t)}$. The causal ordering of the 
model variables (Fig.~\ref{f:uas-dbn}) informs the structure of the \ac{dbn}:
the estimated \ac{cte} and \ac{he} at time $t$ are inputs to the controller 
which, in turn, impacts the future location of the aircraft at time 
$t+\varepsilon$. The directed links between the corresponding variables 
in adjacent time slices capture this dependency. For example, in 
Fig.~\ref{f:uas-dbn}, these are the directed links 
$\estimatedCTE^{(t-1)} \to \actualCTE^{(t)}$, and 
$\estimatedHE^{(t-1)} \to \actualCTE^{(t)}$ (and likewise for the 
preceding and subsequent time slices).
The directed links $\actualCTE^{(t-1)} \to \actualCTE^{(t)}$ model the 
correlation between actual vehicle position over time, also capturing vehicle
inertia. 

At time $t$, the runway image $I^{(t)}$ influences the belief about the true 
aircraft location, \ie the states of $\actualCTE^{(t)}$, with the node $D$ 
modeling the associated structural uncertainty. This reflects the intuition 
that upon detecting an outlier image (more generally an \ac{ood} input), we are 
no longer confident that the image seen is an indicator of the actual aircraft 
location. Fig.~\ref{f:uas-dbn} reflects these dependencies by the directed edges 
$\actualCTE^{(t)} \gets I^{(t)}$, and $\actualCTE^{(t)} \gets D^{(t)}$, respectively. 

Fig.~\ref{f:uas-dbn} shows two adjacent time slices of the \ac{dbn} structure, 
although the actual structure is unrolled for $T$ time steps, the duration of 
taxiing, to compute the assurance measure over the taxi phase. At time $t$, this 
is, in fact, the sum of the probability mass over the seven states of 
$\actualCTE^{(t)}$ that lie within the interval $[-2, 2]$ (see 
Table~\ref{t:uas-model-vars}). 
By unrolling the \ac{dbn} for an additional $\varepsilon$ time steps 
and propagating the uncertainty through the model from the time of the 
last observations, the model can provide an assurance forecast.

\subsubsection{Probability Distributions} 

To complete the \ac{dbn} model specification, we need to specify the 
\acp{cpd} over the model variables, as encoded by its structure. One way to 
identify the \acp{cpd} is through \acl{uq} of the physical system 
model~\cite{najm2009uncertainty}. Practically, the latter may not be 
available, especially for \acp{les}. 

Another alternative---the approach we take here---is to assume a functional form 
for the \acp{cpd} that is then tuned based on execution and simulation data. 
Specifically, to construct the \ac{cpd} represented by the transition edge 
between the time slices, \ie $\prob(\actualCTE^{(t)}\,|\, \actualCTE^{(t-1)}, \estimatedCTE^{(t-1)}, \estimatedHE^{(t-1)})$, we chose a multinomial 
distribution with a uniform prior, tuned using the \ac{map} estimate 
on simulation data. This choice was advantageous in the sense that the 
\ac{dbn} produces a uniform posterior distribution over $\actualCTE$ 
when the observed variables take on values from a distribution different from
that of the data used to build the \acp{cpd}. 
For this example, the simulation data comprised sequences of runway 
images, estimated \ac{cte} and \ac{he} as produced by the \ac{cnn}, and 
true \ac{cte}. Section~\ref{s:uas-experimental-results} gives more details 
on the simulation platform and data gathered. 

To determine the emission probability $\prob(\actualCTE^{(t)}\,|\,I^{(t)})$, 
first we used the \ac{gp} model underpinning our prior work on component-level 
assurance quantification~\cite{adp-edcc2019}. In brief, the idea is to use 
a \ac{gp} to model the error performance of the \ac{cnn} (\ie its accuracy) 
on its input (\ie runway images). Then, adding the error distribution to the 
estimate of \ac{cte} gives the distribution over the true \ac{cte}. However, 
for high dimensional data (such as images), this is computationally expensive. 
Instead, in this paper we used an ensemble of decision
trees~\cite{criminisi2012decision} as a classifier that ascribes a 
probability distribution over the states of $\actualCTE$, given a runway 
image, $I$. This approach builds uncorrelated decision trees such that 
their combined estimate is more accurate than that of any single decision tree. 
To identify the decision rules, we used supervised learning over the 
collection of runway images and corresponding true \ac{cte}, sampled from 
the same environments used to train and test the \ac{cnn} (see 
Section~\ref{s:uas-experimental-results}). For this example, we built $280$ 
decision trees with terminal node size of at least $10$, by randomly sampling 
$100$ data points using the \emph{Gini index} as a performance metric, 
selecting the model parameters to balance classification accuracy and 
computational resources.

\section{Experimental Results}\label{s:uas-experimental-results}

We now present some results of our experiments in quantifying \ac{les} 
assurance in terms of the assurance measure, $\prob\left(|\actualCTE| < 
2\meter\right)$, based upon simulations of constant speed taxiing missions. 

\subsection{Simulation Setup}\label{ss:uas-simulation-setup}

We use a commercial-off-the-shelf flight simulator instrumented to reflect 
the pipeline architecture of Fig.~\ref{f:uas-example}. The simulation 
environment includes various airports and runways with centerlines of 
varying quality, \eg portions of the centerline may be obscured at 
various locations (see Fig.~\ref{f:uas-example}). 
We can create different training and test environments by changing various 
simulation settings, among which two that we have selected are: 
\begin{inparaenum}[\itshape i\upshape)]
	\item weather induced visibility (\emph{clear} and \emph{overcast}), 
	and 
	\item the time of day (\emph{07:30am} to \emph{2:00pm}).
\end{inparaenum} 	
Two such environments are, for example, ``\emph{Clear at 07:30am}'', 
and ``\emph{Overcast at 12:15pm}''. More generally, we can construct 
environments such as ``\emph{Clear Morning}'', ``\emph{Overcast Afternoon}'', 
and so on. The former refers to the collection of data sampled from the 
environment having clear weather, and the time of day incremented in steps 
of $15$ and $30$ minutes from \emph{07:30am} until \emph{noon}. A similar 
interpretation applies to other such environments. 

From these environments, we gathered images via automated screen capture 
(simulating the camera output) whilst taxiing the aircraft on the airport 
runway, using different software controllers, as well as different \acp{cnn} 
for perception: \ie the same \ac{cnn} architecture described in 
Section~\ref{s:example}, but trained by our industrial collaborators with 
data drawn from the various environments identified earlier. In tandem, for 
each image, we collected true \ac{cte} (from internal simulation variables), 
along with estimates of \ac{cte} and \ac{he}. 
We used several such data sets, one for each of the different 
environments identified above, from which data samples were drawn to build
the \acp{cpd} of the \ac{dbn} model. Here, note that these data samples 
were \emph{not} identical to those used to train and test the \ac{cnn}, 
even though the samples were drawn from the collection of environments 
common to both the \ac{les} and the \ac{dbn}.

\begin{figure}[t]
	\centering
	\includegraphics[width=0.75\columnwidth]{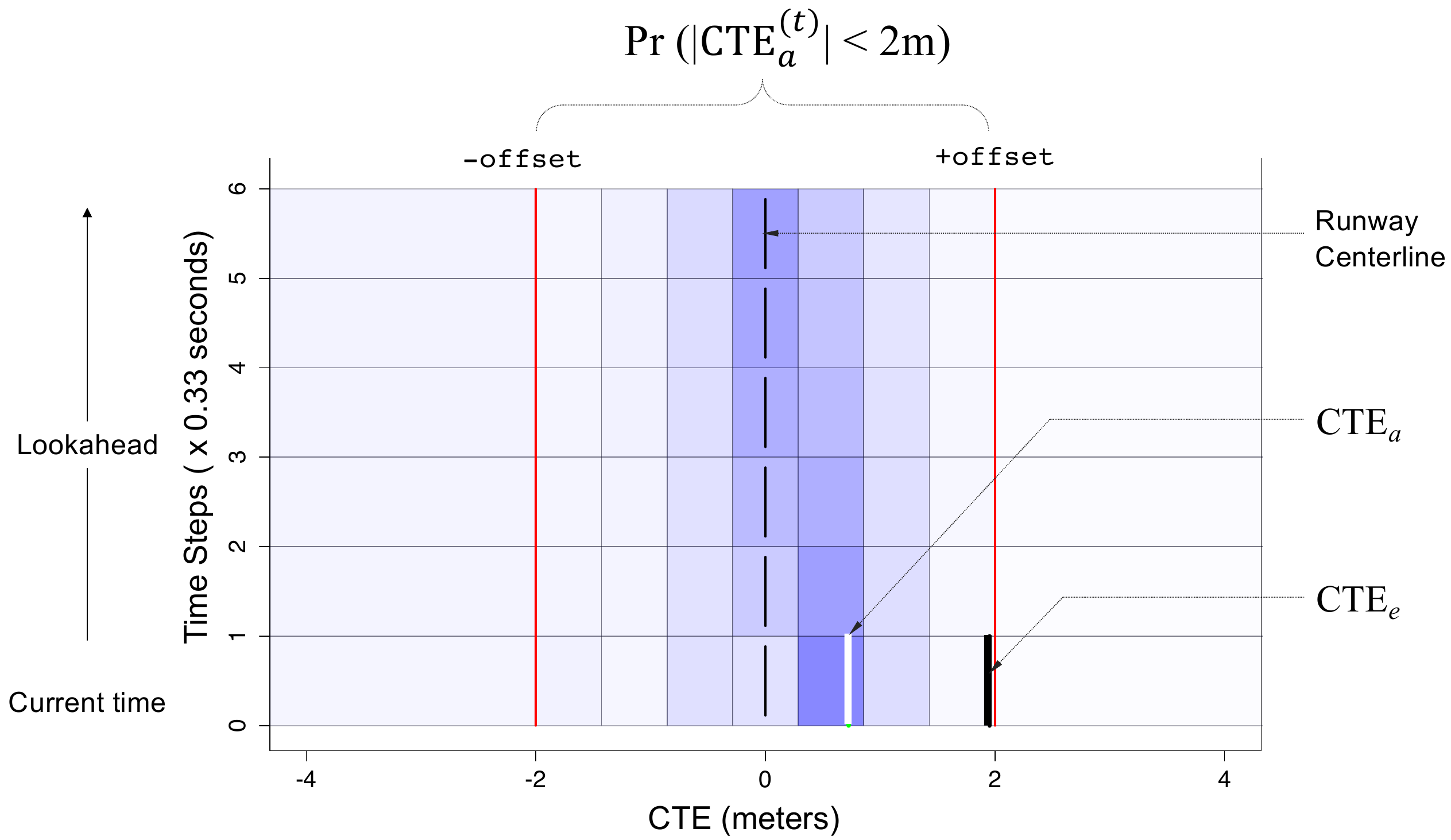}
	\caption{Visualization of predicted uncertainty in true \acl{cte}, 
		$\actualCTE^{(t)}$, to quantify assurance in runway centerline tracking 
		as the assurance measure, $\prob(\assuredTaxi)$.} 
	\label{f:uas-uncertainty}{}
	\vspace{\baselineskip}
	\includegraphics[width=0.725\columnwidth]{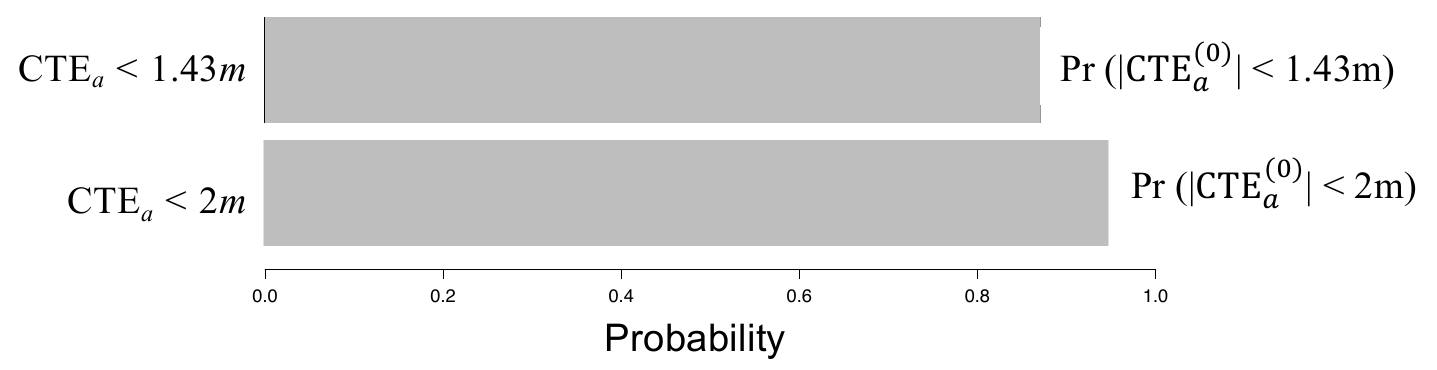}
	\caption{$\prob(\assuredTaxi)$ for $\offset=2\meter$ 
		and $\offset = 1.43\meter$.}
	\label{f:uas-assurancemeasure}
\end{figure}

\subsection{Uncertainty Quantification}

Fig.~\ref{f:uas-uncertainty} shows the results of assurance quantification for 
one test scenario, visualized as a probability surface overlaid on a stretch 
of the runway, itself shown as a grid. The horizontal axis---discretized using 
the interval boundaries for the states of $\actualCTE$ (see 
Table~\ref{t:uas-model-vars})---gives the true aircraft location, which is 
uncertain during taxiing. Thus, moving from left to right (or vice versa) 
constitutes lateral aircraft movement. 
The vertical axis (discretized into $6$ steps, each of duration 
$0.33\second$) represents the number of time slices for which the \ac{dbn} 
model is unrolled. We selected this based on the time taken for the \ac{uas} 
to laterally depart the runway after violating the $2\meter$ bound, given: 
runway dimensions, maximum allowed taxiing speed, and other constraints on 
the \ac{uas} dynamics, \eg non-accelerating taxiing. 

At $t=0$, the horizontal axis gives the aircraft location at the current time. 
The time steps $t=1,\ldots, 6$ are \emph{lookahead times} for which the 
horizontal axis gives the \emph{predicted} location of the aircraft relative 
to the centerline, given the \ac{cnn} estimates of \ac{cte} and \ac{he} at $t=0$. 
Thus, moving from the bottom to the top of Fig.~\ref{f:uas-uncertainty} 
represents forward taxiing, \ie the temporal evolution of aircraft position
over the runway. 
Each cell of the grid formed by discretizing the two axes is, therefore,
a state of $\actualCTE$ at a given time, shaded such that darker shades 
indicate lower uncertainty (or higher confidence) and lighter shades 
indicate higher uncertainty (or lower confidence). 
Thus, the row at $t=0$ shows the \ac{dbn} estimate of uncertainty over 
$\actualCTE$ at the current time. Similarly, each row for $t=1, \ldots, 5$ 
shows the \emph{predicted} uncertainty over $\actualCTE$ for those 
lookahead times, given that the last known values for the observed 
variables are at $t=0$.
The solid white line in Fig.~\ref{f:uas-uncertainty} at $t=0$ is 
\emph{ground truth}, \ie the true \ac{cte} at the current time based on 
internal simulation variables. Although this may not be otherwise 
available during taxiing, we show it here primarily for model validation, 
\ie to show that the interval (state of $\actualCTE$) estimated 
by the \ac{dbn} to be the least uncertain is also the one that includes the 
ground truth. 
The solid black line is \ac{cte} as estimated by the \ac{cnn} 
(\ie $\estimatedCTE$) at the current time. 

Recall that assured taxiing involves maintaining $\actualCTE$ between 
a $2\meter$ lateral offset on either side of the centerline. To quantify 
assurance in this property, we sum up the probability mass in each cell 
between the two offsets. Fig.~\ref{f:uas-assurancemeasure}
shows the assurance measure, $\prob(|\actualCTE^{(t=0)}| < \offset)$ computed 
for two different offset values: $2\meter$ and $1.43\meter$.\footnote{The 
introduction of a second offset was motivated by our industry collaborators 
to integrate the assurance measure on the \ac{les} platform.} 
The interval $[-2, 2]$ is a Bayesian \emph{credible interval} 
within which the true \ac{cte} lies with probability $\approx 95\%$, 
based on Fig.~\ref{f:uas-assurancemeasure}. In other words, \emph{the 
\ac{dbn} model is $\approx 95\%$ confident that the aircraft is truly 
located within $2\meter$ of the runway centerline}.
In general, the expected (and desired) \ac{dbn} behavior is to be more 
uncertain over longer term assurance forecasts, when there are no 
additional observations with which to update the posterior distributions 
on the assurance measures.

\subsection{Sufficient Assurance} 

We must select a threshold on the assurance measure to establish what 
sufficient assurance constitutes, based on which we can assert whether or 
not the assurance claim holds. The criterion we have selected here is: when 
the \ac{dbn} is $\geq 30\%$ confident that the true \ac{uas} location exceeds 
the allowed lateral offset, the assurance claim does not hold, \ie 
$\Pr(|\actualCTE^{(t)}| \geq 2\meter) \geq 0.3 \Rightarrow \neg(\assuredTaxi)$.
We determined this threshold under conservative assumptions about vehicle 
behavior, leveraging the engineering judgment of our industry collaborators, 
to balance the tradeoff between safety (avoiding runway overrun) and mission 
effectiveness (not stopping too often).

\section{Discussion}\label{s:discussion}

We now evaluate how the \ac{dbn} performs relative to the \ac{les}, in 
the context of ground truth. 
The intent is to show that it is a reasonable (\ie valid) \emph{reference model} 
of the system suitable for runtime use (\ie simple and abstract), based on which 
to make certain decisions, \eg whether or not to stop taxiing. Moreover, we 
must also show that the software implementation of the \ac{dbn} can be 
relied upon. In this paper, we primarily address the former, leaving the latter 
for future work. 

\subsection{Validity}

We compare how well the \ac{dbn} and the \ac{les} can discriminate
between \emph{true positive} and \emph{true negative} situations when their 
respective outputs are transformed into a classification on a plurality of 
image data drawn from multiple simulated taxiing scenarios for different test
environments unseen by both the \ac{dbn} and the \ac{les}.  

A true positive (negative) situation for the \ac{dbn} is one where it indicates 
that the assurance property is satisfied (not satisfied) based on the criterion
for sufficient assurance (see Section~\ref{s:uas-experimental-results}), 
and ground truth data also indicates that it is truly the case that the \ac{uas}
location is within (exceeds) the allowed lateral offset from the runway 
centerline. 
Likewise for the \ac{les}, a true negative (positive) situation is one where
the \ac{cnn} estimate of \ac{cte} indicates (does not indicate) an offset 
violation \ie $\estimatedCTE \geq 2\meter$ (equivalently, $\estimatedCTE < 
2\meter$), and so does ground truth data. 

\begin{table}[ht]
	\centering
	\caption{\ac{dbn} Performance evaluation for runway centerline tracking.}
	\label{t:uas-model-validation}
	\includegraphics[width=0.9\columnwidth]{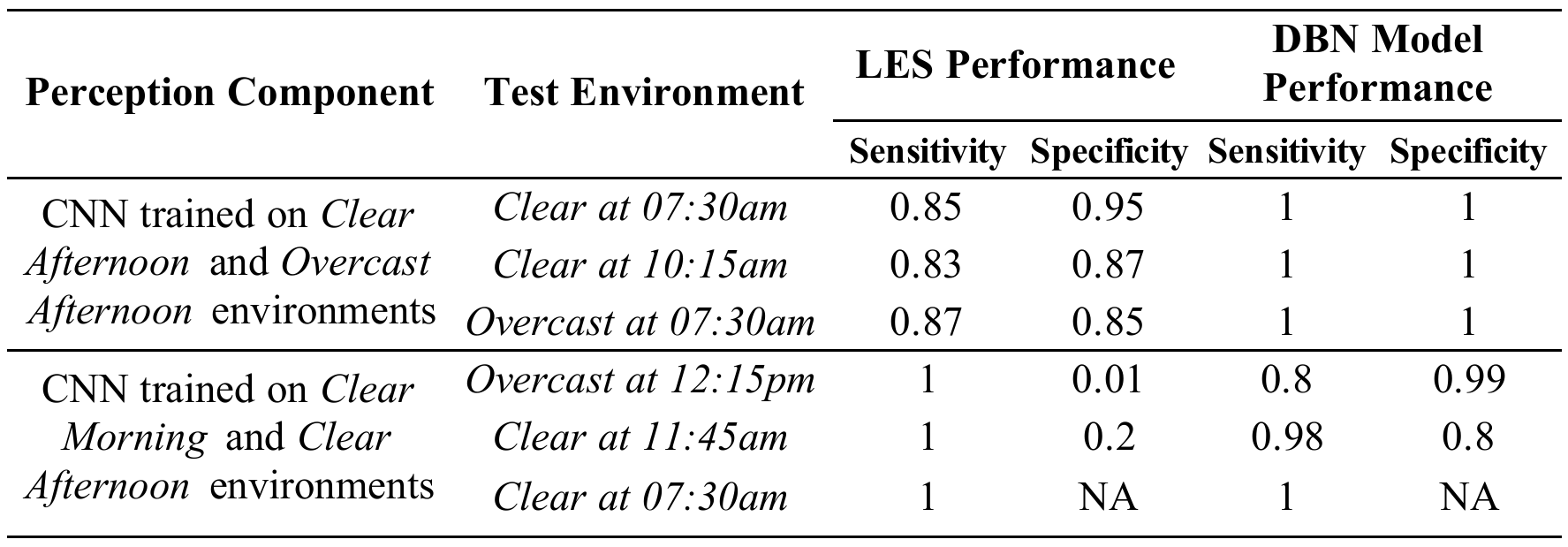}
\end{table}

Table~\ref{t:uas-model-validation} shows our evaluation results in terms of 
\emph{sensitivity} (true positive rate) and the \emph{specificity} (true 
negative rate) of both the \ac{dbn} model and the \ac{les}, varying the 
embedded \ac{cnn} used for perception. The variability arises from using 
\acp{cnn} trained under two different training environments. We also used 
these training environments to build the \ac{dbn} for both \ac{les} variants 
using $\approx 37000$ image samples. These samples were not the same as those 
that were used to train the \ac{cnn} variants: indeed, we did not have access 
to the actual training data for the different \acp{cnn}.
Also, the test environments listed in the table (and, therefore, the resulting 
test data), are unseen during the development of both \ac{les} variants, 
and the \ac{dbn} models of the same.

Based on Table~\ref{t:uas-model-validation}, in the context of the sensitivity 
and specificity metrics shown, as well as the criterion for sufficient 
assurance, we are cautiously optimistic in claiming that the \ac{dbn} 
models the \ac{les} reasonably well. 
For the test environments ``\emph{Clear at 11:45am}'', and ``\emph{Overcast at 
12:15pm}'', the \ac{dbn} has a lower sensitivity than the \ac{les}, however its
specificity is substantially better. This suggests that the \ac{les} may be 
biased in its estimates of \ac{cte} for those operating conditions.

\subsection{Suitability} 

The \ac{dbn} model structure---in particular, the conditional independence 
relations encoded by the structure---is informed by (our knowledge of) the 
causal impacts of the identified variables and the system dynamics, and the 
resulting assumptions. 
We note that it is always possible to relax these assumptions and learn the 
\ac{dbn} structure as well as its parameters. However, in most cases, 
especially when there is limited data available, structure learning can be 
an unidentifiable problem, or can produce a non-unique solution. 
In our case, the conditional independence assumptions used have turned out 
to be neither too strong to affect model performance nor too conservative to 
impose a problem in identifying the \acp{cpd} given limited data.

Our assessment in Table~\ref{t:uas-model-validation} does \emph{not}
compare the \ac{dbn} and the \ac{cnn} that estimates \ac{cte}. 
Indeed, the latter is a learned, static regression function for a 
\emph{component}, that associates a vector of real values with a real-valued 
scalar, whereas here we are assessing a stochastic process model of a 
(learning-eabled) \emph{system} (\ie the \ac{dbn}) against the system itself. 
When we use the \ac{dbn} for runtime assurance, we implement it as a software
component integrated into the \ac{les}. This can be viewed as an item to which
we can apply our own assurance methodology, \ie as in Section~\ref{s:approach}, 
and~\cite{adp-edcc2019}. Thus, although we have not formulated assurance 
properties for the \ac{dbn}, sensitivity and specificity are probabilistic 
performance metrics (albeit in a frequentist sense) that we can view as
assurance measures in their own right, that we have now applied to our model.

The validation above is admittedly not exhaustive although the following 
observations are worth noting: the \ac{dbn} is a relatively simple and 
abstract model of the time-series evolution of the \emph{system}, whose 
estimates can be updated through Bayesian inference given observed data. 
Thus, it is amenable to applying other verification techniques including 
inspection, and formal verification. 

Moreover, the \ac{dbn} \emph{does not} produce point estimates of \ac{cte}; 
rather, in quantifying confidence in a system-level assurance property, a 
by-product is the uncertainty in true \ac{cte} given as a probability 
distribution over the range of admissible values of $\actualCTE$. 
Thus, in unseen situations where the \ac{cnn} can produce an inaccurate 
estimate of \ac{cte} (see Fig.~\ref{f:uas-uncertainty}), the \ac{dbn} 
gives a distribution over possible values of true \ac{cte}. As sucsh, it is 
more conservative in potentially unsafe scenarios. Based on this assessment, 
we submit that the \ac{dbn} is a reasonable and suitable runtime reference 
model of the \ac{les} for the autonomous taxiing application, when used for 
centerline tracking.

\acused{cms}

\subsection{Utility} 

A key advantage of an abstract assurance quantification model is a small 
implementation footprint for runtime integration into the \ac{les}.
As indicated in Section~\ref{s:intro}, one of the primary motivations for 
quantified assurance measures is to provide feedback signals 
(in a computable form) to the \ac{les}, that can be acted on, \eg by a \emph{\acl{cms}} (\acs{cms}), in operation.
In this work, the assurance measure values were translated into commands 
to either \emph{stop}, \emph{slow down}, or \emph{continue} based on 
\begin{inparaenum}[\itshape i\upshape)]
	\item the chosen decision thresholds 
	(Section~\ref{s:uas-experimental-results}), and 
	\item a simple model of the system-level effect (\ie likelihood of lateral
	runway overrun) given the assurance measure and current system state.%
	\footnote{Although the content of integrating assurance measures with
	a \ac{cms} is very closely related to the work here, it is not in 
	scope for this paper, and will be the topic of a forthcoming article.}
\end{inparaenum}
In general, deciding between a series of options in the presence of 
conflicting and uncertain outcomes is a special case of \emph{decision 
making under uncertainty}~\cite{Kochenderfer2015}. We plan to 
investigate such techniques as future work to develop a principled approach 
to contingency management using assurance measures. 

The aim of \emph{run-time assurance}, also known as \emph{run-time
verification}, is to provide updates as to whether a system satisfies
specified properties as it executes \cite{R2U2}. This is done using a run-time
\emph{monitor}, which evaluates the property using values extracted from the
state of the system and its environment. In a sense, therefore, the notion of
assurance measure we have described here is a kind of monitor. However, it is
worth making several distinctions. A monitor relates directly to properties of
the system, whereas an assurance measure characterizes \emph{confidence} in
our knowledge of such properties. Second, an assurance measure seeks to
aggregate  a range of sources of information, including monitors. Thus it can
be seen as a form of \emph{data fusion}. Third, monitors typically provide
values that relate to the current state of the system, whereas the assurance
measures we have defined are predictive, intended to give a probabilistic 
quantification on dependability attributes. 

In general, our approach to assurance quantification admits other models 
including runtime monitors: recall that the node $D^{(t)}$ in 
Fig.~\ref{f:uas-dbn} is a runtime monitor detecting data distribution 
shift in the input image at time $t$. 
Indeed, our framework is not intended to replace runtime verification, 
and the assurance measures generated show the assurance contribution of the 
runtime monitors, additionally providing an assurance/uncertainty forecast. We 
are not aware of existing runtime verification techniques that do this.

\section{Related Work}\label{s:related-work}

The work in this paper is closely related to our earlier research on 
assurance case confidence quantification~\cite{dhp-esem-11}. There, 
although confidence estimation in an assurance claim also uses Bayesian 
techniques, it relies primarily on the argument structure to build the 
model. Similarly, based on the structure of an argument, the use of an 
evidential theory basis has been explored for confidence quantification 
in assurance claims~\cite{guiochet2019}. However, neither work has been 
applied to \ac{les} assurance quantification. Moreover, in this paper 
the focus is on those properties where quantification is possible, relying 
upon models of the system that can be assessed against objective, measured 
data. 

This paper is a natural extension of our prior work on quantifying assurance 
in \ac{ml} components~\cite{adp-edcc2019}: the assurance property we consider 
there is $\estimatedCTE$ \emph{accuracy}. Assurance quantification then entails 
using \acfp{gp} to determine the uncertainty in the error of $\estimatedCTE$,
which is inversely proportional to accuracy. However, the data used are not 
(and need not be) time dependent and the model used applies regardless of 
whether or not the aircraft position has violated $\assuredTaxi$. Indeed, 
despite a high assurance \ac{cnn} that accurately estimates \ac{cte}, it is 
nevertheless possible to violate $\assuredTaxi$. However, in this paper we 
model the \ac{les} as a stochastic process, including any runtime 
mitigations, \eg a monitor for detection \ac{ood} images. As such, the 
models used for \ac{uq} are a generalization of that in~\cite{adp-edcc2019}
to time-series behavior. 

As previously indicated (Section~\ref{s:intro}), one of the motivations is to 
support dynamic \acfp{ac}. Our prior work~\cite{dhp-icse2015} first explored 
this concept, which has subsequently been tailored for so-called 
\emph{self-adaptive software}~\cite{calinescu2017}. Again, neither work has 
considered \acp{les}, although self-adaptation is one of the properties that 
\acp{les} can exhibit.
In~\cite{dhp-icse2015}, confidence quantification has been situated as a 
core principle of dynamic assurance which has also motivated this paper to 
an appreciable degree. However, that work relies on the quantification 
methodology in~\cite{dhp-esem-11}. 
In \cite{calinescu2017}, assurance quantification employs probabilistic
model checking, which can be leveraged for \acp{les} if they can be represented
using state-space models, \eg as in~\cite{verisig} which uses hybrid model
checking instead. Neither technique is incompatible with the stochastic
processes-based modeling approach that we have adopted. As such, they may be 
a candidate means to check properties of the stochastic models that we build 
as a means of (meta-)assurance. 

\emph{Dynamic safety management} as an assurance concept has also been proposed
as a run-time assurance method~\cite{trapp2018}, but it is largely speculative 
about applicability for \acp{les}.
The idea of \emph{requirements-aware} runtime models~\cite{bencomo2019} is 
very closely related to our notion of building a reference model. Quantified 
and probabilistic guarantees in reinforcement learning have been explored in 
developing assured \ac{ml} components in~\cite{kochenderfer2019-rl}. That 
work is also closely related to what we have presented here, though its focus 
is mainly on assurance of correctness properties that have a safety impact. 
Additionally, the assurance approach there is \emph{intrusive} in the sense 
that the \ac{ml} component being built is modified. In our case, assurance 
quantification does not modify the \ac{ml} components. 
\emph{Benchmarking} of uncertainty estimation 
techniques~\cite{weiss-uncertainty-benchmarking} has also been investigated, 
although mainly in the context of image classification. It is unclear if the 
reported results translate to assurance quantification as applied in this paper. 
However, the benchmarking principles and metrics used could be candidates for 
evaluating various system models built using our approach. 

Kalman filters have long been used to address uncertainty during state
estimation, and have some similarities to our approach. A Kalman filters is a 
special case of a \ac{dbn} where amongst the main assumptions are that sensor 
errors are distributed as zero mean Gaussians, and that the uncertainty does 
not vary between sensing outputs. In contrast, our model uses discrete 
distributions, admitting varying sensor uncertainty for each image input, in a 
more general graphical model that has a different structure, whilst including 
detections of \ac{ood} inputs.

\section{Conclusion and Future Work}\label{s:conclusions}

We have described our approach to quantifiable assurance using assurance 
measures, run-time computations of uncertainty (conversely, confidence) 
in specified assurance properties, and their application to \acfp{les}.
Assurance measures complement design-time assurance activities, each of which
forms part of an overall \acf{dac}.
In collaboration with system integrators from industry, we have applied 
our framework to an aviation platform that employed supervised learning 
using a deep \ac{cnn}. Collaboration was crucial to develop the 
contingency management capability, which relied on engineering judgment 
to tradeoff safety risk reduction and achieving performance objectives. 
Feedback from the end-users (\ie our industry collaborators) was also essential 
in refining the final visualizations of the assurance measure that we 
ultimately deployed in the system (based on Fig.~\ref{f:uas-assurancemeasure}). 
Those are intended to provide insight into the system assurance state 
for safety observer crew.

We have shown that our methodology can feasibly quantify assurance in
system-level properties of an aviation domain \ac{les}, though we have 
used classical \ac{uq} techniques. 
Our work in quantifying assurance in \acp{les} is ongoing, and we will be 
developing assurance measures for other autonomous platforms in the context of 
more complex mission objectives that require additional \ac{ml} components and 
learning schemes. 

The work in this paper is one strand of our overall approach to assurance through \acp{dac}. The diverse components of an assurance case, including structured 
arguments, safety architecture~\cite{dpw-ress2019}, as well as the assurance 
measures described here, each represent one facet of an integrated \ac{dac}.
There are close connections between the probabilistic models underlying
assurance measures and the safety architecture, as well as between assurance
properties and claims in an assurance arguments. Our future work will
place these connections on a rigorous basis. In part, this can be achieved
through use of a high-level \emph{\acl{dsl}} (\acs{dsl}) that will let us 
\begin{inparaenum}[\itshape i\upshape)]
	\item abstract from the details of the individual probabilistic models, and
	\item conversely, allow compilation into a range of different models, 
	whilst making more explicit the connections to domain concepts used 
	elsewhere in the assurance case.
\end{inparaenum}

A related avenue of future work is providing comprehensive assurance 
for our approach itself, and in turn, the assurance measures produced. From a 
verification standpoint, we can consider \emph{correctness} properties 
entailing
\begin{inparaenum}[\itshape i\upshape)]
	\item consistency between the quantification model and the other \ac{dac} 
	components, \eg the risk scenarios captured by a safety architecture, and
	\item correctness of the low-level implementation against the higher level
	specification embodied by the quantification model. 
\end{inparaenum} 
Additionally, assurance measure validity is related, in part, to the limits
of the statistical techniques used to infer the underpinning stochastic models, 
and the data used to build them.

Indeed, one of the challenges we faced in this work was obtaining sufficient 
useful data. Moreover, the quality of the data gathered also plays a key role 
in corroborating that the assurance quantification models sufficiently represent
the system behavior across its intended operational profile. We believe that a 
more principled approach to specifying a variety of training data should be 
possible (\eg to include various types of perturbed and adversarial inputs), 
and that such specifications could be derived from the \acs{dsl} used to specify 
the assurance measures themselves.
The dynamic nature of \acfp{ac} will also bear further investigation,
to see how real-time updates provided by assurance measures during a mission
can inform updates between missions, to the qualitative arguments of \acp{ac}.

%
%
\bibliographystyle{ieee}

\end{document}